\documentclass[aps,prd,nofootinbib,amsmath,amssymb,superscriptaddress,twocolumn,10pt]{revtex4}
\usepackage{amsmath}
\usepackage{amsfonts}
\usepackage{txfonts}
\usepackage{graphicx}
\usepackage{dcolumn}
\usepackage{bbm}
\usepackage{amssymb}
\usepackage{latexsym}
\usepackage{CJK}
\usepackage{titlesec}
\usepackage[colorlinks=true, linkcolor=red, citecolor=blue]{hyperref}
\begin{document}
 \bibliographystyle{unsrt}

\title{Statefinder hierarchy exploration of the extended Ricci dark energy}

\author{Fei Yu\footnote{Corresponding author}}
\email{yufei@sau.edu.cn}
\affiliation{College of Sciences, Shenyang Aerospace University, Shenyang 110136, China}
\author{Jing-Lei Cui}
\affiliation{Department of Physics, College of Sciences, Northeastern University, Shenyang 110004, China}
\author{Jing-Fei Zhang}
\affiliation{Department of Physics, College of Sciences, Northeastern University, Shenyang 110004, China}
\author{Xin Zhang}
\affiliation{Department of Physics, College of Sciences, Northeastern University, Shenyang 110004, China}
\affiliation{Center for High Energy Physics, Peking University, Beijing 100080, China}

\begin{abstract}
We apply the statefinder hierarchy plus the fractional growth parameter to explore the extended Ricci dark energy (ERDE) model, in which there are two independent coefficients $\alpha$ and $\beta$. By adjusting them, we plot evolution trajectories of some typical parameters, including the Hubble expansion rate $E$, deceleration parameter $q$, the third- and fourth-order hierarchy $S_3^{(1)}$ and $S_4^{(1)}$ and fractional growth parameter $\epsilon$, respectively, as well as several combinations of them. For the case of variable $\alpha$ and constant $\beta$, in the low-redshift region the evolution trajectories of $E$ are in high degeneracy and that of $q$ separate somewhat. However, the $\Lambda$CDM model is confounded with ERDE in both of these cases. $S_3^{(1)}$ and $S_4^{(1)}$, especially the former, perform much better. They can differentiate well only varieties of cases within ERDE except $\Lambda$CDM in the low-redshift region. For the high-redshift region, combinations $\{S_n^{(1)},\epsilon\}$ can break the degeneracy. Both of $\{S_3^{(1)},\epsilon\}$ and $\{S_4^{(1)},\epsilon\}$ have the ability to discriminate ERDE with $\alpha=1$ from $\Lambda$CDM, of which the degeneracy cannot be broken by all the before-mentioned parameters. For the case of variable $\beta$ and constant $\alpha$, $S_3^{(1)}(z)$ and $S_4^{(1)}(z)$ can only discriminate ERDE from $\Lambda$CDM. Nothing but pairs $\{S_3^{(1)},\epsilon\}$ and $\{S_4^{(1)},\epsilon\}$ can discriminate not only within ERDE but also ERDE from $\Lambda$CDM. Finally, we find that $S_3^{(1)}$ is surprisingly a better choice to discriminate within ERDE itself, and ERDE from $\Lambda$CDM as well, rather than $S_4^{(1)}$.
\end{abstract}
\maketitle

\renewcommand{\thesection}{\arabic{section}}
\renewcommand{\thesubsection}{\arabic{subsection}}
\titleformat*{\section}{\flushleft\bf}
\titleformat*{\subsection}{\flushleft\bf}

\section{Introduction}
Data from a series of astronomical observations for more than a decade have shown that the universe is undergoing an epoch of accelerated expansion~\cite{Riess:1998AnJ1009,Perlmutter:1999ApJ565,Spergel:2007ApJS377,Adelman:2008ApJS297}. The most likely explanation for this cosmic acceleration is that the universe is currently being dominated by an exotic component, named {\it dark energy} (DE), which exerts repulsive gravity. To explain the origin and physical properties of dark energy, numerous theoretical/phenomenological models have been proposed~\cite{Bamba:2012ASS155}. Among the models, the most successful one is the $\Lambda$CDM model (which mainly includes the cosmological constant $\Lambda$ and cold dark matter), because it is simple but could provide a very good fit to the observational data currently available. The cosmological constant is equivalent to the vacuum energy density with $w=-1$. For a time-dependent equation-of-state parameter (EOS) $w$, there are lots of models, such as quintessence~\cite{Steinhardt:1999PRD123504}, Chaplygin gas~\cite{Kamenshchik:2001PLB265}, holographic dark energy~\cite{LM:2004PLB1}, and so on.

In this paper, we study the model inspired by the holographic principle of quantum gravity. The holographic principle was enlightened by quantum properties of black hole~\cite{Bekenstein:1973PRD2333,Bousso:1999JHEP07004} and later extended to string theory~\cite{Susskind:1994JMP6377}. According to the work of Cohen et al.~\cite{Cohen:1999PRL4971}, when $\rho_{\rm de}$ is taken as the quantum zero-point energy density caused by a short distance cut-off, the total energy in a region of size $L$ should not be more than the mass of a black hole of the same size, i.e., $L^3\rho_{\rm de} \leqslant LM_{\rm p}^2$. The saturated form of this inequality, which is equivalent to the largest $L$ allowed, leads to the energy density of the holographic dark energy, $\rho_{\rm de}=3c^2M_{\rm p}^2L^{-2}$, where $c$ is a numerical constant introduced and $M_{\rm p}$ is the reduced Planck mass with $M_{\rm p}^2=(8\pi G)^{-1}$. For the model setting, the choice of the infrared (IR) cut-off $L$ is very crucial. After the denial of the Hubble scale~\cite{Hsu:2004PLB13} and the particle horizon~\cite{Bousso:1999JHEP07004,Fischler:9806039} as IR cut-off for their failure to give rise to the cosmic acceleration, Li chose the future event horizon instead, getting the expected success~\cite{LM:2004PLB1}. But the adoption of the future event horizon indicates that the history of dark energy depends on the future evolution of the scale factor $a(t)$, which violates causality~\cite{CRG:2007PLB228}. Then the agegraphic dark energy model~\cite{CRG:2007PLB228,WH:2008PLB113} and the Ricci dark energy (RDE) model~\cite{GCJ:2009PRD043511} emerged to avoid the violation of causality. The former is characterized by the age of the universe as the length measure while the latter takes the average radius of Ricci scalar curvature $|R|^{-1/2}$ as the IR cut-off. Further, the RDE model was extended to a more general form, liberating coefficients of the two terms, of the energy density~\cite{Nojiri:2006GRG1285,Granda:2008PLB275}
\begin{equation}\label{erde}
\rho_{\rm de}=3M_{\rm p}^2(\alpha H^2+\beta\dot{H}),
\end{equation}
where $\alpha$ and $\beta$ are constants to be determined and the dot denotes a derivative with respect to time. For the Ricci-type holographic DE models are determined by a local concept of Ricci scalar curvature rather than a global one of future event horizon, they are naturally free of the causality problem. The model is called the {\it extended Ricci dark energy} (ERDE) model, with the special case $\alpha=2\beta$ the RDE model.

With the increasing number of DE models, diagnostics aiming to differentiate them are needed. So far several methods have appeared. They are the well-known statefinder~\cite{Sahni:2003JETPL201,Alam:2003MNRAS1057}, $Om$~\cite{Sahni:2008PRD103502} and growth rate of perturbations~\cite{Acquaviva:2008PRD043514,Acquaviva:2010PRD082001,WLM:1998ApJ483}. The statefinder is a sensitive and robust geometrical diagnostic of DE, which uses both the second and the third derivatives of $a(t)$. Recently, Arabsalmani and Sahni further extended the statefinder to higher-order derivatives of $a(t)$, and called such a diagnostic ``statefinder hierarchy''~\cite{Arabsalmani:2011PRD043501}. The statefinder diagnostic has been applied to various DE models~\cite{ZX:2005PLB1,ZX:2005IJMPD1597,ZX:2006JCAP01003,Setare:2007JCAP03007,ZJF:2008PLB26,FCJ:2008PLB231,TML:2009PRD023503,ZL:2010IJMPD21,YF:2013CTP243,CJL:2014EPJC2849,Sahni:2014ApJL40}, but sometimes we do need the diagnostic with higher-order derivatives of $a(t)$. For instance, when diagnosing the new agegraphic DE model, the original statefinder (second and third derivatives) cannot differentiate this model with different parameter values~\cite{CJL:2014EPJC2849}, but the hierarchy (further higher-order derivatives) is capable of breaking the degeneracy~\cite{CJL:2014EPJC3100}.

Here we study the ERDE model with statefinder hierarchy, supplemented by the growth rate of perturbations, to explore what the behaviors are like when ERDE takes different parameter values and what the difference is
between ERDE and $\Lambda$CDM. Necessarily referring to, in our previous work~\cite{YF:2013CTP243}, we have diagnosed, with the original statefinders, the ERDE model both with interaction between DE and matter and not. The results therein seems satisfactory since there is no appearance of degeneracy for ERDE with various parameter values. But from the aspect of completeness of a theory, we neglected another evolution tendency of ERDE, mentioned in some papers~\cite{ZX:2009PRD103509,Granda:09100778,Mathew:2013IJMPD1350056}, that $\alpha$ of values larger than 1 enables the ERDE model to exhibit another orientation of evolution symmetrical to that plotted in Ref.~\cite{YF:2013CTP243}. As a matter of fact~\cite{Granda:09100778}, $\alpha>1$ makes ERDE behave like quintessence~\cite{Steinhardt:1999PRD123504} ($w>-1$), while $\alpha<1$ like quintom~\cite{FB:2005PLB35} ($w$ evolves across the cosmological-constant boundary $w=-1$). We will expatiate on this theme in a later section.

In Sect.~2, the ERDE model is exhibited. In Sect.~3, we introduce the diagnostic tools of statefinder hierarchy and growth rate of perturbations. Then ERDE will be explored in Sect.~4. Finally Sect.~5 gives the conclusion.

\section{The ERDE model}
We consider a flat universe with DE and matter, namely,
\begin{equation}\label{fdm}
3M_{\rm p}^2H^2=\rho_{\rm de}+\rho_{\rm m},
\end{equation}
where $\rho_{\rm de}$ and $\rho_{\rm m}$ are, respectively, energy densities of DE and matter, and $\rho_{\rm de}$ takes the form of ERDE described by Eq.~(\ref{erde}).

We get
\begin{equation}\label{eq4}
E^2=\frac{H^2}{H_0^2}=\Omega_{\rm m0}e^{-3x}+\alpha E^2+\frac{\beta}{2}\frac{dE^2}{dx},
\end{equation}
where $E=H/H_0$ is the dimensionless Hubble expansion rate, $x=\ln a$ and the subscript `` 0 '' denotes present values of physical quantities. The solution of Eq.~(\ref{eq4}) is
\begin{equation}\label{eq7}
E^2=\Omega_{\rm m0}e^{-3x}+\frac{3\beta-2\alpha}{2\alpha-3\beta-2}\Omega_{\rm m0}e^{-3x}+f_0e^{\frac{2}{\beta}(1-\alpha)x},
\end{equation}
where
\begin{equation}
f_0=1+\frac{2}{2\alpha-3\beta-2}\Omega_{\rm m0},
\end{equation}
under the initial condition $E_0=E(x=0)=1$.

The fractional density and EOS of ERDE are given by
\begin{eqnarray}
\Omega_{\rm de} &=& \frac{1}{E^2}\frac{\rho_{\rm de}}{\rho_0} \nonumber \\
&=& \frac{1}{E^2}\left(\frac{3\beta-2\alpha}{2\alpha-3\beta-2}\Omega_{\rm m0}e^{-3x}+f_0e^{\frac{2}{\beta}(1-\alpha)x}\right),\label{omegade} \\
w &=& \frac{\frac{2\alpha-3\beta-2}{3\beta}f_0e^{\frac{2}{\beta}(1-\alpha)x}}{\frac{3\beta-2\alpha}{2\alpha-3\beta-2}\Omega_{\rm m0}e^{-3x}+f_0e^{\frac{2}{\beta}(1-\alpha)x}}.\label{w}
\end{eqnarray}

\section{Statefinder hierarchy and growth rate of matter perturbations}

\subsection*{3.1 The statefinder hierarchy}
The primary aim of the statefinder hierarchy is to single the $\Lambda$CDM model out from evolving DE ones~\cite{Arabsalmani:2011PRD043501}. For it has a convenient property that all members of the statefinder hierarchy can be expressed in terms of some elementary functions (like the deceleration parameter $q$, the EOS $w$ or the fractional density $\Omega$), even the Chaplygin gas, interacting dark energy, and modified gravity models have already been explored in this way~\cite{LJ:2014JCAP12043,YL:150308948,Myrzakulov:2013JCAP10047}.

To review briefly, we just explain the primary principle of the statefinder hierarchy. Because in Ref.~\cite{Arabsalmani:2011PRD043501}, the EOS of DE $w$ in the hierarchy expressions is a constant, we here generalize $w$ to be time-dependent and have intensive interest only in the final expressions of statefinder hierarchy members in terms of elementary functions. Later we will see that these elementary functions are the fractional density and EOS of ERDE, already derived above, Eqs.~(\ref{omegade}) and (\ref{w}).

We Taylor-expand the scale factor $a(t)$ around the present epoch $t_0$:
\begin{equation}
\frac{a(t)}{a_0}=1+\sum\limits_{n=1}^{\infty}\frac{A_n(t_0)}{n!}[H_0(t-t_0)]^n,
\end{equation}
where
\begin{equation}
A_n=\frac{a^{(n)}}{aH^n},~~~n\in N;
\end{equation}
$a^{(n)}$ is the $n$th derivative of $a(t)$ with respect to time. The familiar term $A_2=-q$ represents the deceleration parameter, while $A_3$ is the very original statefinder ``$r$''~\cite{Sahni:2003JETPL201}. $A_4$ was ever referred to as the snap ``$s$''~\cite{Visser:2004CQG2603}, while $A_5$ the lerk ``$l$''~\cite{Dabrowski:2005PLB184}. For $\Lambda$CDM ($w=-1$),
\begin{equation}\label{eq13}
\begin{split}
&A_2=1-\frac{3}{2}\Omega_{\rm m},\\
&A_3=1,\\
&A_4=1-\frac{3^2}{2}\Omega_{\rm m},~~~~~{\rm etc},
\end{split}
\end{equation}
where $\Omega_{\rm m}=\frac{2}{3}(1+q)$, which means that for $\Lambda$CDM the elementary functions are the deceleration parameter $q$ or the fractional density parameter of matter $\Omega_{\rm m}$, because $w$ is constant. Then the {\em statefinder hierarchy} $S_n$ can be defined as~\cite{Arabsalmani:2011PRD043501}:
\begin{equation}\label{eq14}
\begin{split}
&S_2=A_2+\frac{3}{2}\Omega_{\rm m},\\
&S_3=A_3,\\
&S_4=A_4+\frac{3^2}{2}\Omega_{\rm m},~~~~~{\rm etc}.
\end{split}
\end{equation}
Comparing Eq.~(\ref{eq14}) with Eq.~(\ref{eq13}), one can get the essential feature of this diagnostic that all the $S_n$ parameters stays pegged at unity for $\Lambda$CDM during the entire course of cosmic expansion,
\begin{equation}
S_n|_{\Lambda\rm{CDM}}=1.
\end{equation}
In fact, that is why $S_n$ are defined in this way, to differ from both other constant-$w$ DE models and evolving ones.

Remember that in Ref.~\cite{Sahni:2003JETPL201} there is a statefinder pair $\{r,s\}$, where $r$ is $S_3$ and $s \equiv \frac{r-1}{3(q-1/2)}$. $s$ also belongs to the third derivative hierarchy and serves the aim of breaking some of the degeneracy present in $r$. To normalize letters of the alphabet, Arabsalmani and Sahni introduced a general pair $\{S_n^{(1)},S_n^{(2)}\}$~\cite{Arabsalmani:2011PRD043501}
\begin{equation}
\begin{split}
&S_3^{(1)}=A_3,\\
&S_4^{(1)}=A_4+3(1+q),~~~~~{\rm etc},
\end{split}
\end{equation}
and
\begin{equation}
S_n^{(2)}=\frac{S_n^{(1)}-1}{\gamma\left(q-\frac{1}{2}\right)},
\end{equation}
where $\gamma$ is an arbitrary constant and the superscript ``(1)''  is used for discriminating not only between the original hierarchy $S_n$ and $S_n^{(1)}$, but also between $S_n^{(1)}$ and its derivative $S_n^{(2)}$. Therefore, $\{S_n^{(1)},S_n^{(2)}\}=\{1,0\}$ for $\Lambda$CDM and $\{r,s\}$ is just $\{S_3^{(1)},S_3^{(2)}\}$ with $\gamma=3$. In this paper we use only the $S_n^{(1)}$ series as follows:

\begin{align}
q= &\frac{1}{2}+\frac{3}{2}w\Omega_{\rm de}, \label{q} \\
S_3^{(1)}= &1+\frac{9}{2}\Omega_{\rm de}w(1+w)-\frac{3}{2}\Omega_{\rm de}w', \label{s31} \\
S_4^{(1)}= &1-\frac{27}{2}\Omega_{\rm de}w(1+w)\left(\frac{7}{6}+w\right)-\frac{27}{4}\Omega_{\rm de}^2w^2(1+w) \nonumber \\
&+\frac{3}{2}\Omega_{\rm de}\left[\left(\frac{13}{2}+9w+\frac{3}{2}w\Omega_{\rm de}\right)w'-w''\right], \label{s41}
\end{align}
where the prime denotes the derivative with respect to $x=\ln a$.

\begin{figure}[htbp]
\centering
\includegraphics[scale=0.3]{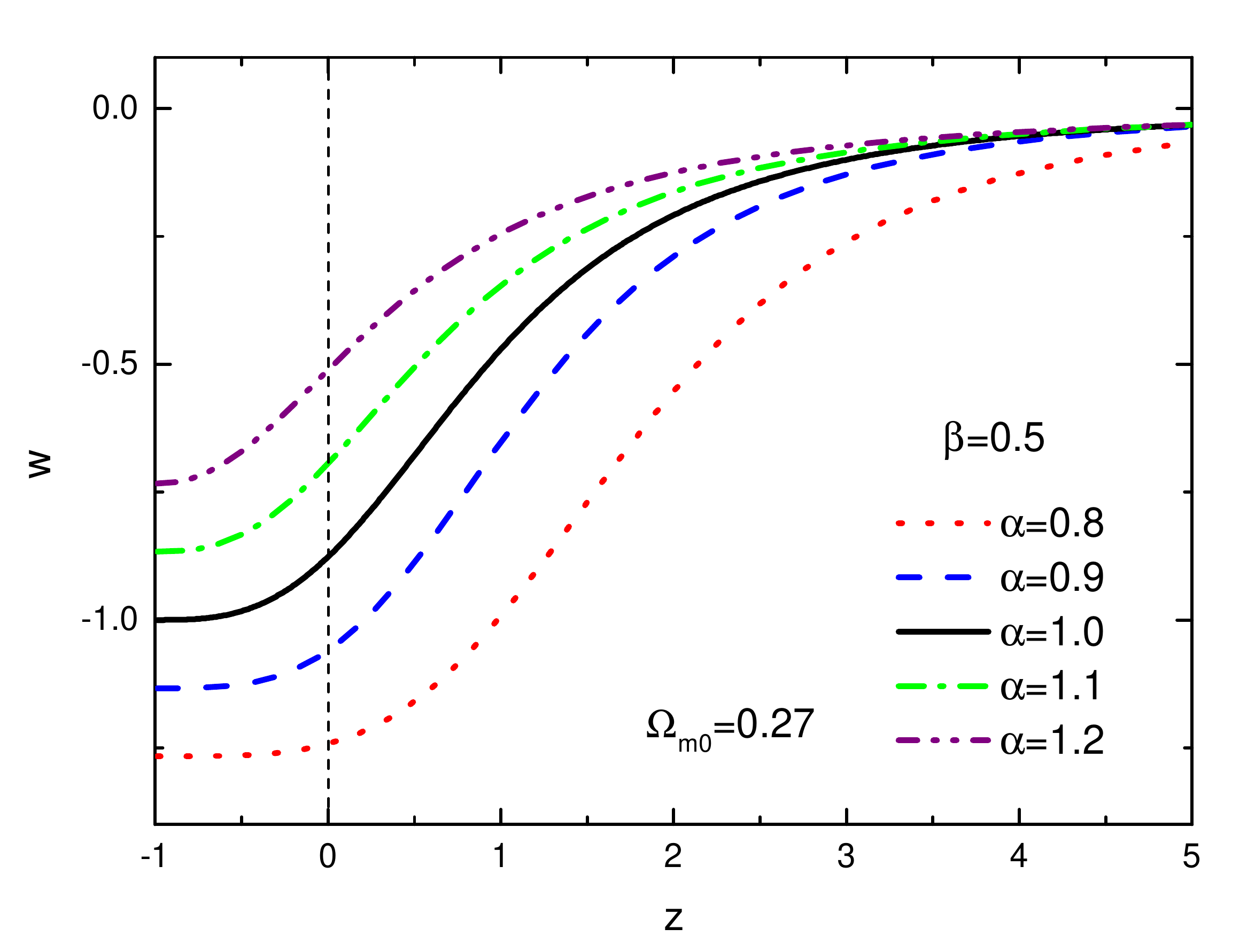}
\caption{(color online). The evolution trajectories of the equation of state $w$ versus redshift $z$ of ERDE for variable $\alpha$ with $\beta=0.5$. Herein $\Omega_{\rm m0}=0.27$.}
\label{fig1}
\end{figure}

\begin{figure*}[htbp]
\centering
\includegraphics[scale=0.3]{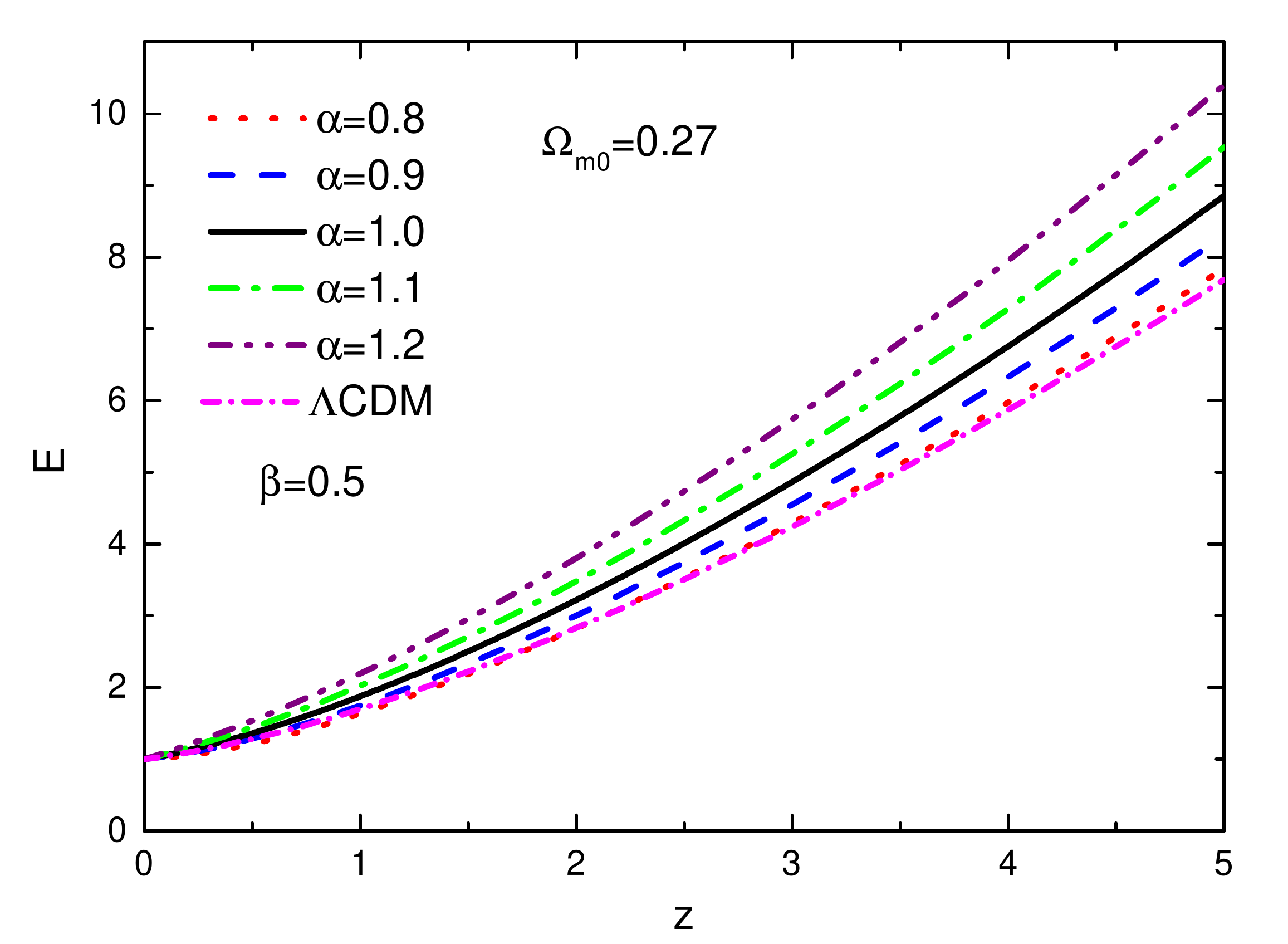}
\includegraphics[scale=0.3]{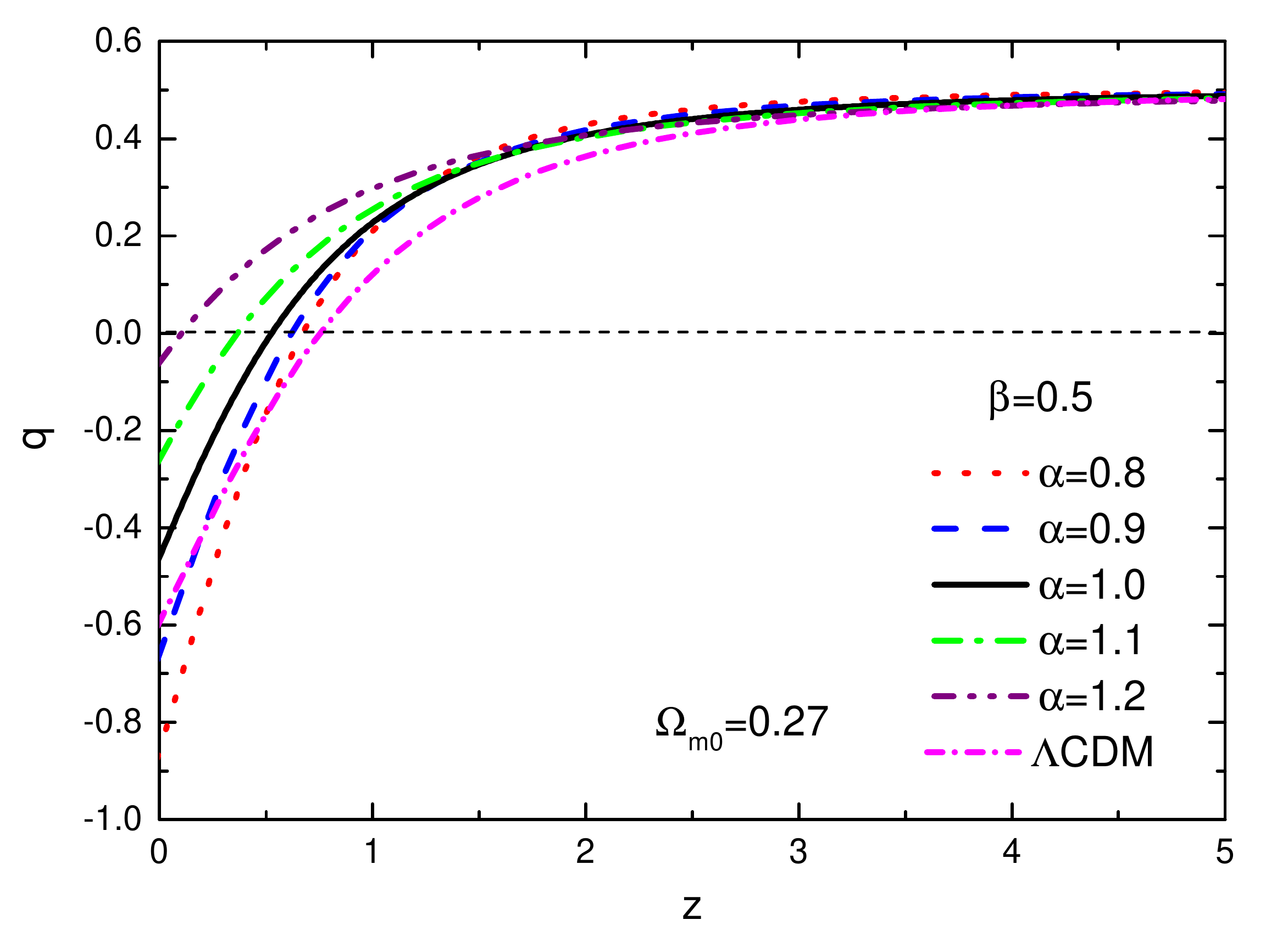}
\includegraphics[scale=0.3]{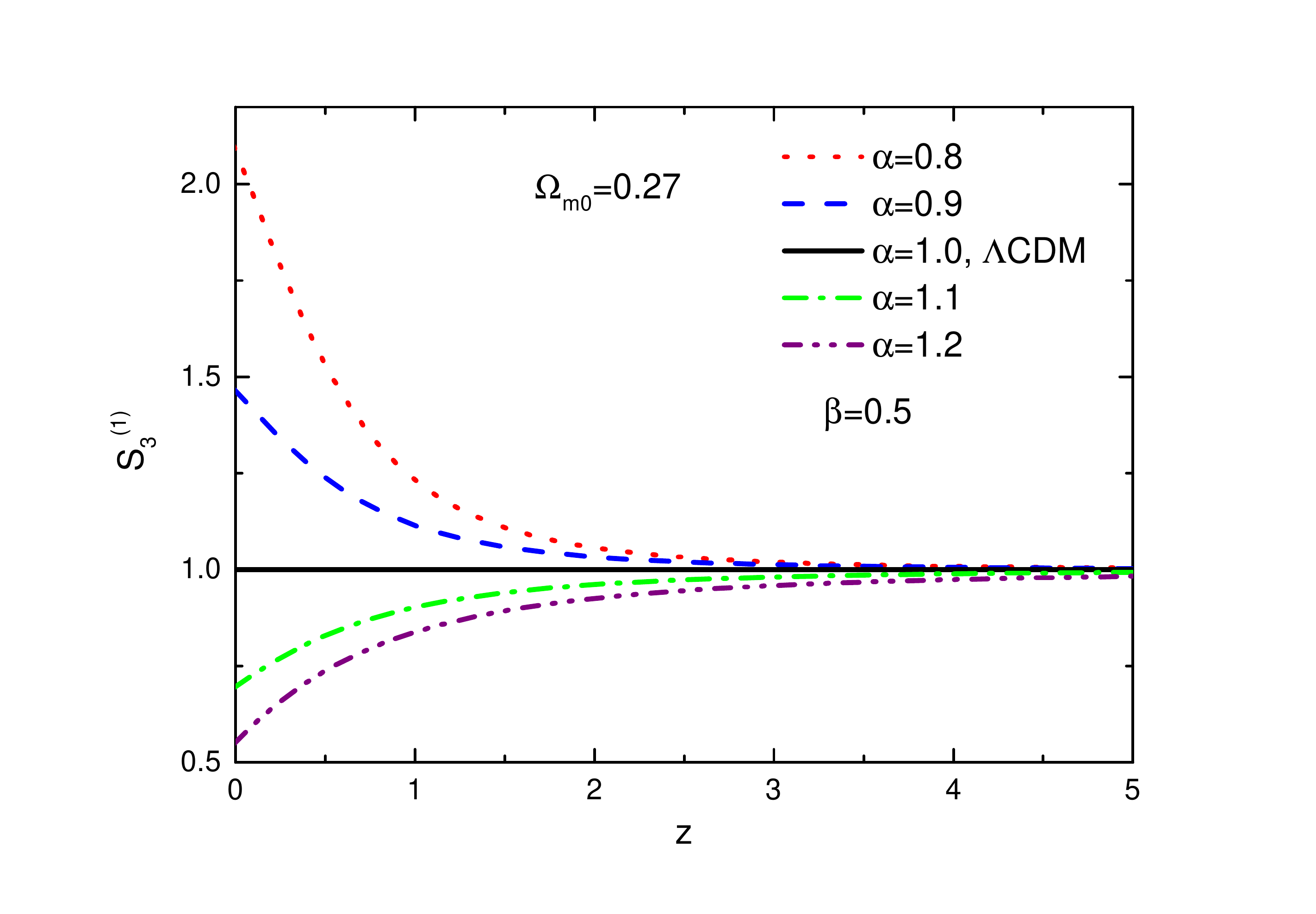}
\includegraphics[scale=0.3]{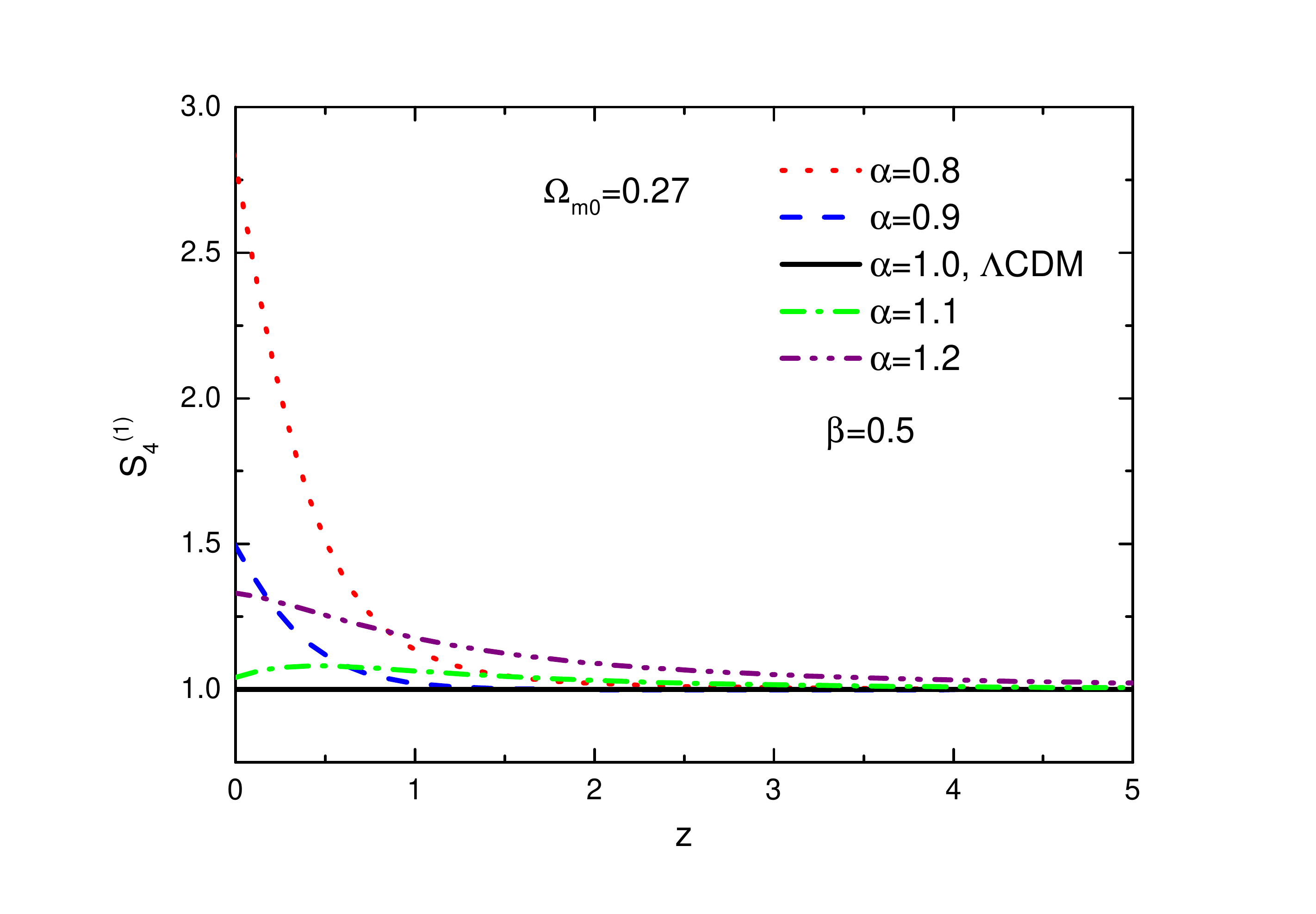}
\caption{(color online). The evolution trajectories of $E$, $q$, $S_3^{(1)}$ and $S_4^{(1)}$, respectively, versus redshift $z$ of ERDE for variable $\alpha$ with $\beta=0.5$, as well compared with the $\Lambda$CDM model. Herein $\Omega_{\rm m0}=0.27$.}
\label{fig2}
\end{figure*}

\subsection*{3.2 The growth rate of perturbations}
The fractional growth parameter $\epsilon(z)$~\cite{Acquaviva:2008PRD043514,Acquaviva:2010PRD082001} can supplement the statefinders as a null diagnostic as well, defined as
\begin{equation}
\epsilon(z):=\frac{f(z)}{f_{\Lambda{\rm CDM}}(z)},
\end{equation}
where $f(z)=d\ln\delta/d\ln a$ represents the growth rate of linearized density perturbations~\cite{WLM:1998ApJ483},
\begin{equation}
f(z) \simeq \Omega_{\rm m}^\gamma(z),
\end{equation}
\begin{equation}
\gamma(z)=\frac{3}{5-\frac{w}{1-w}}+\frac{3(1-w)\left(1-\frac{3}{2}w\right)}{125\left(1-\frac{6}{5}w\right)^3}(1-\Omega_{\rm m}),
\end{equation}
where $w$ either is a constant, or varies slowly with time. Combine the fractional growth parameter $\epsilon(z)$ with statefinder hierarchy to define a {\em composite null diagnostic} (CND): $\{S_n,\epsilon\}$~\cite{Arabsalmani:2011PRD043501}. For $\Lambda$CDM, $\gamma\simeq0.55$ and $\epsilon=1$~\cite{WLM:1998ApJ483,Linder:2005PRD043529}, therefore, $\{S_n,\epsilon\}=\{1,1\}$.

\section{Exploring ERDE with statefinder hierarchy}

\begin{figure*}[htbp]
\centering
\includegraphics[scale=0.3]{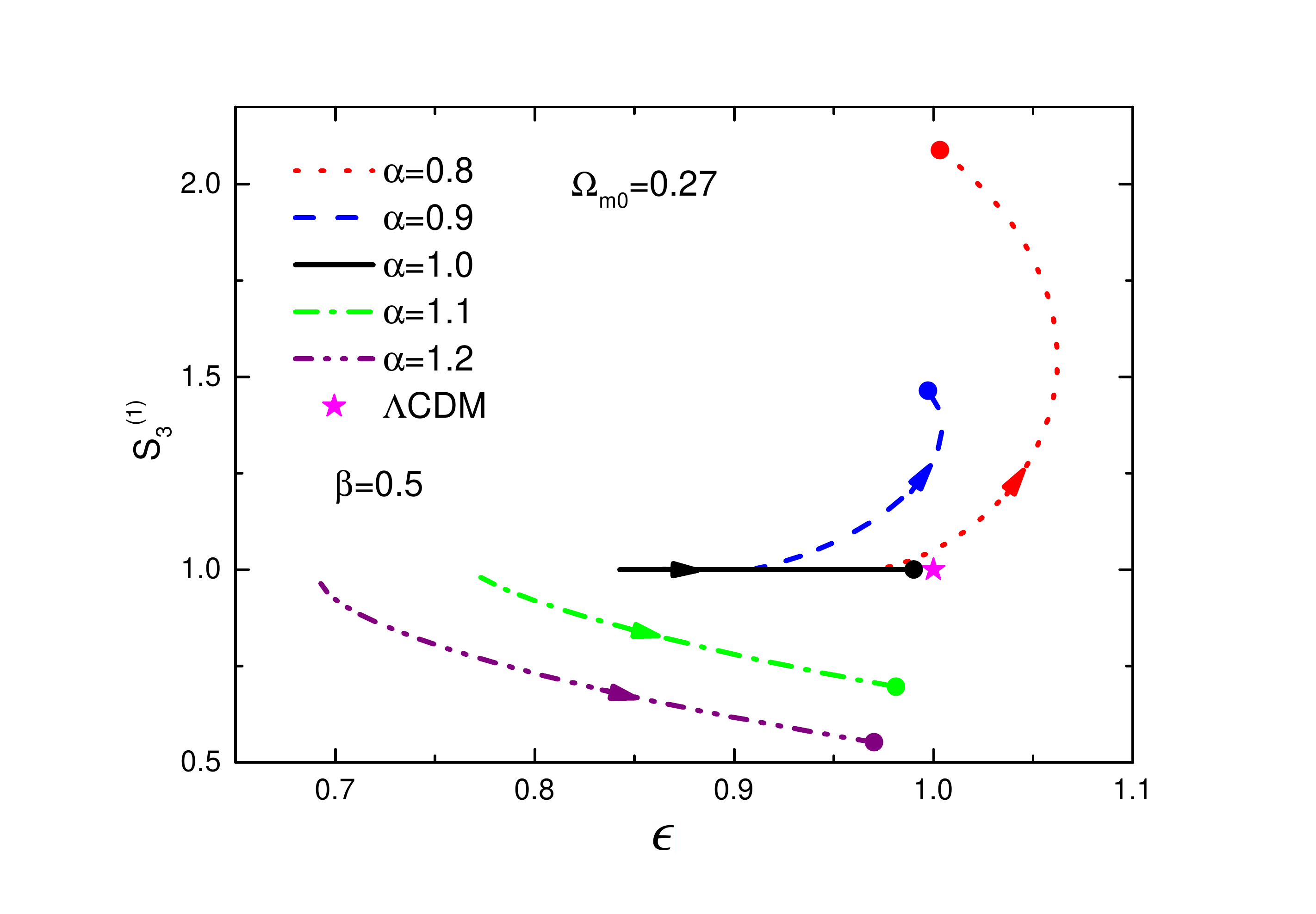}
\includegraphics[scale=0.3]{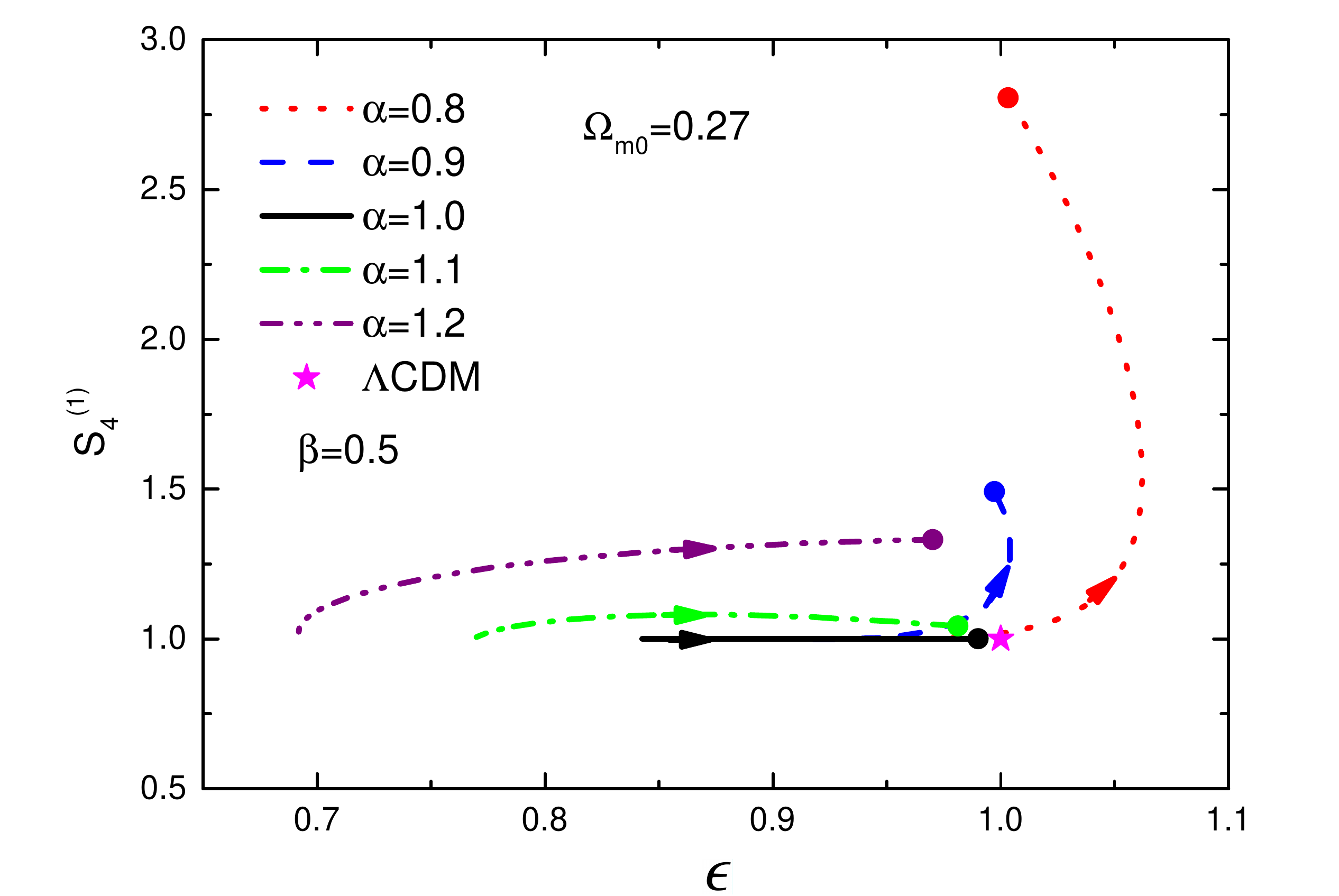}
\caption{(color online). The evolution trajectories of $S_3^{(1)}$ and $S_4^{(1)}$, respectively, versus $\epsilon$ of ERDE for variable $\alpha$ with $\beta=0.5$, as well compared with the $\Lambda$CDM model marked by a star. The dots represent the present values and the arrows indicate the directions of evolution. Herein $\Omega_{\rm m0}=0.27$.}
\label{fig3}
\end{figure*}

\begin{figure}[htbp]
\centering
\includegraphics[scale=0.3]{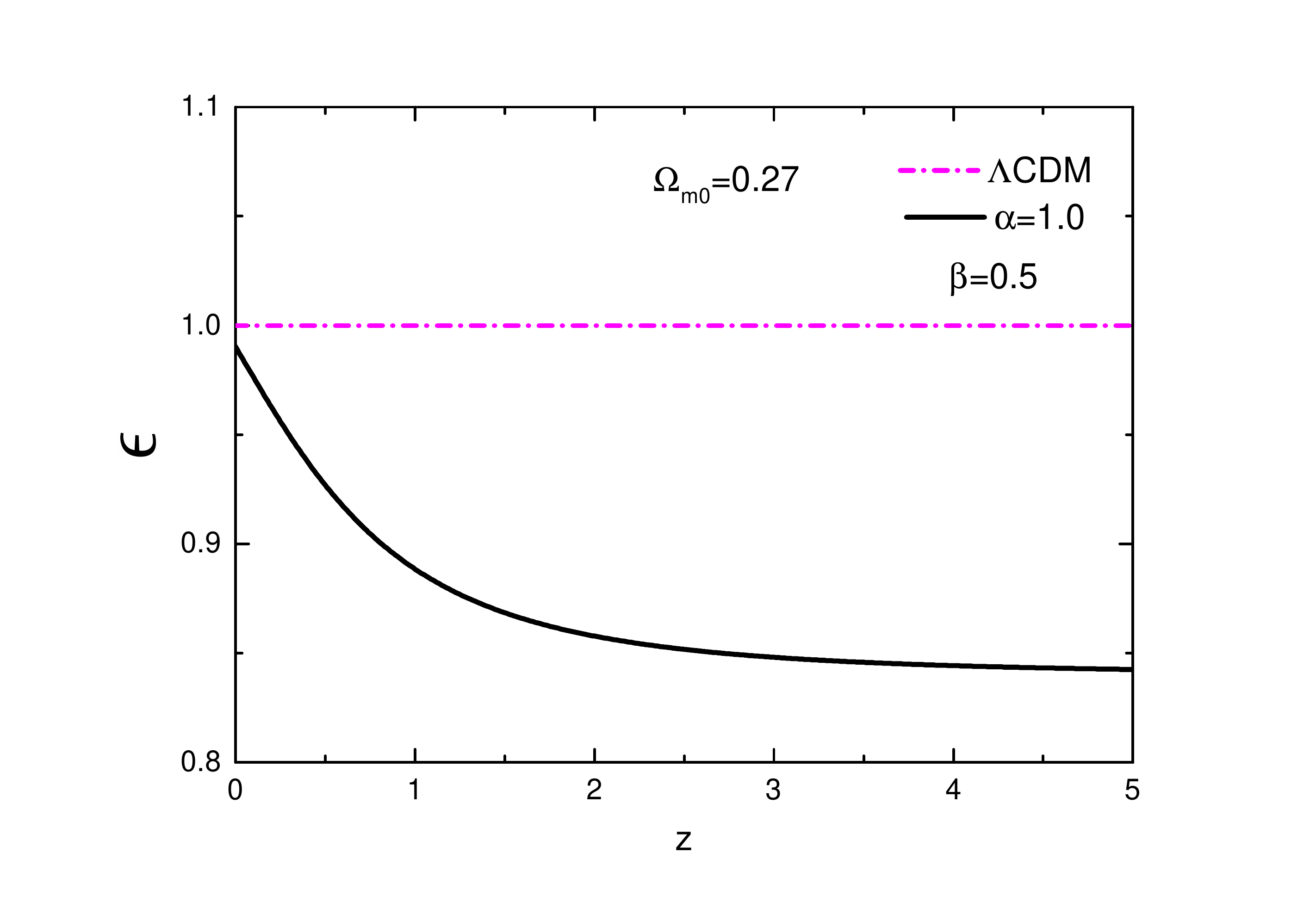}
\caption{(color online). A comparison of evolution trajectories of the fractional growth parameter $\epsilon$ versus redshift $z$ of ERDE for $\alpha=1$ with $\beta=0.5$ and that of the $\Lambda$CDM model. Herein $\Omega_{\rm m0}=0.27$.}
\label{fig4}
\end{figure}

\begin{figure*}[htbp]
\centering
\includegraphics[scale=0.3]{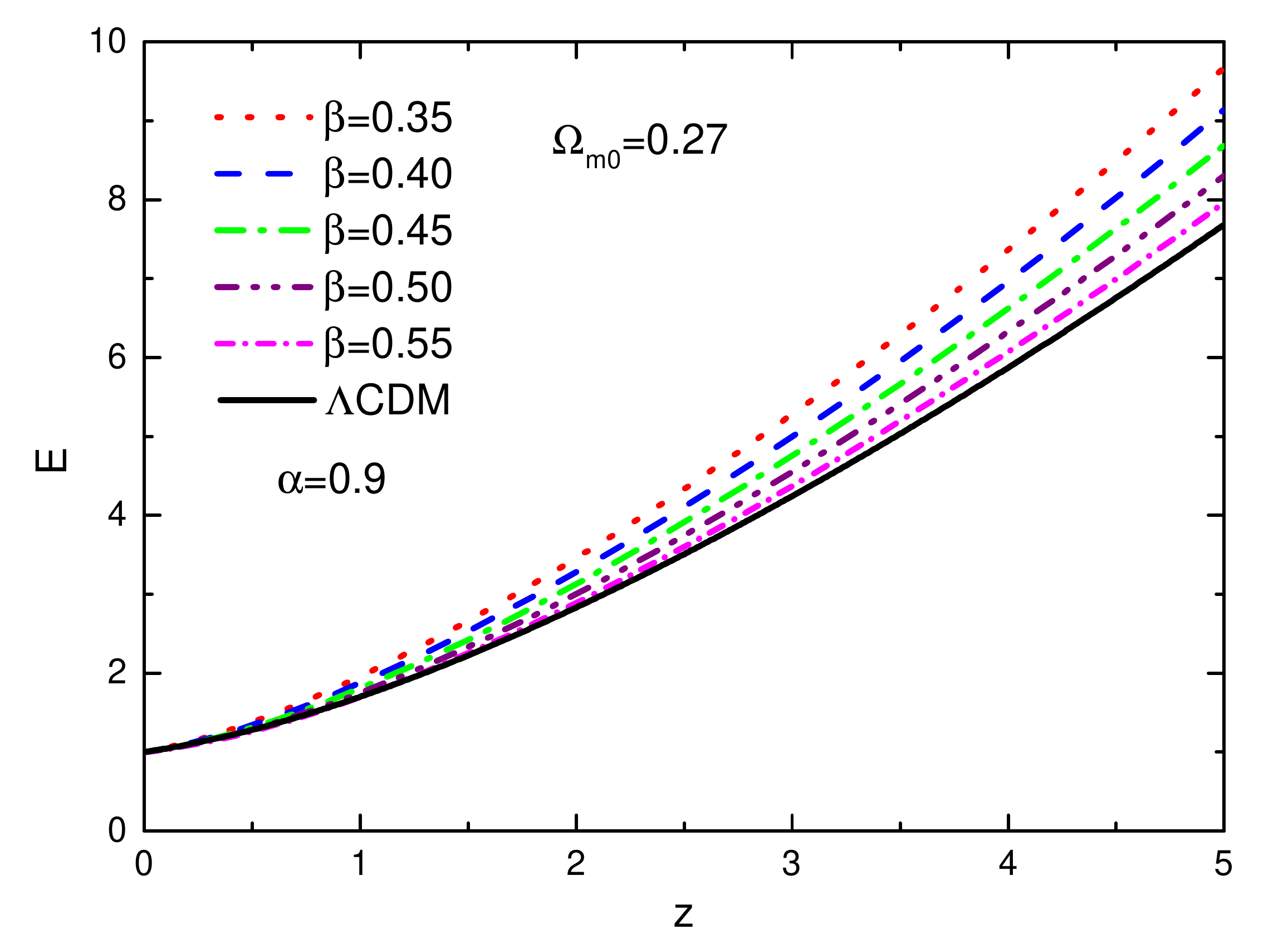}
\includegraphics[scale=0.3]{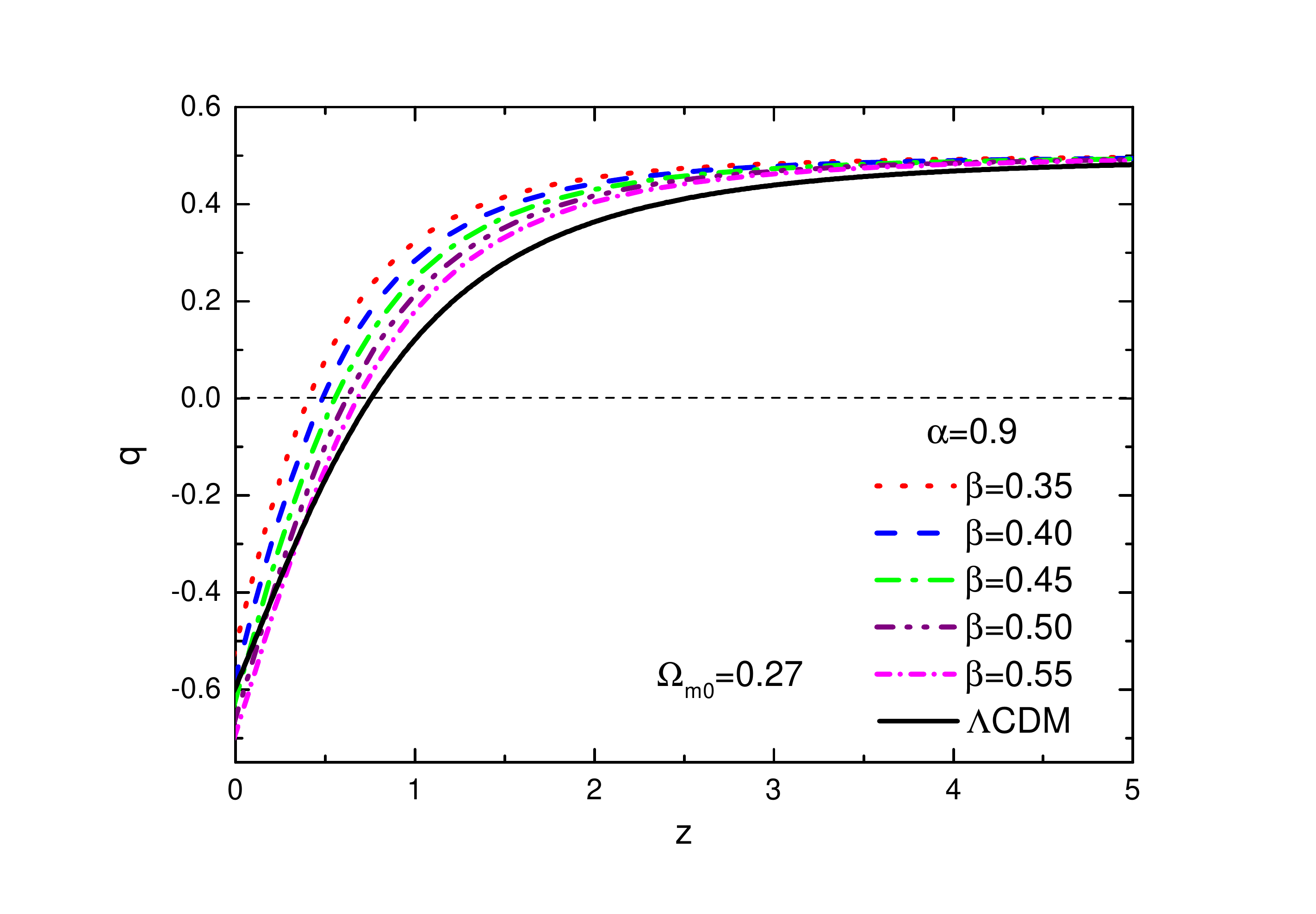}
\includegraphics[scale=0.3]{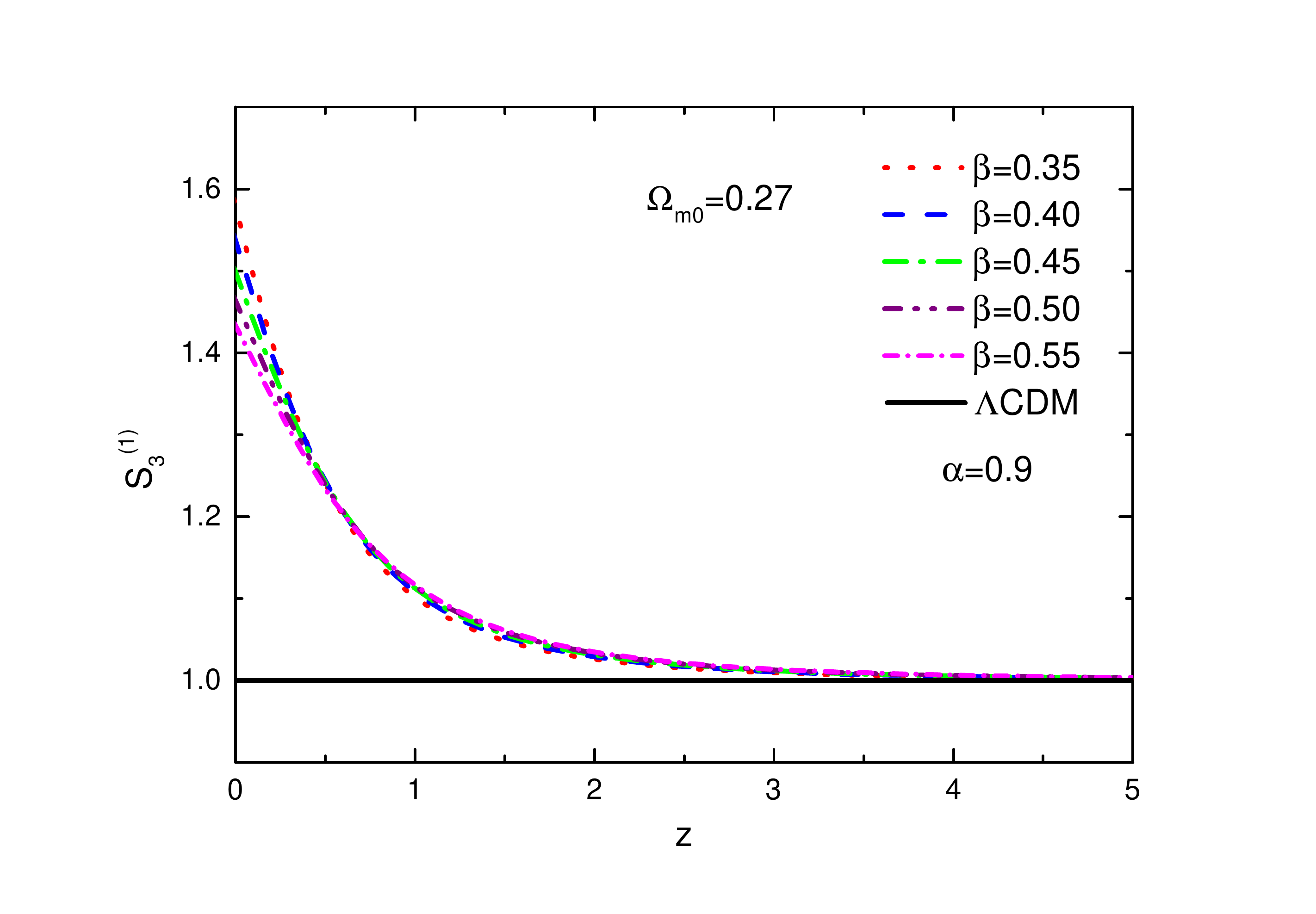}
\includegraphics[scale=0.3]{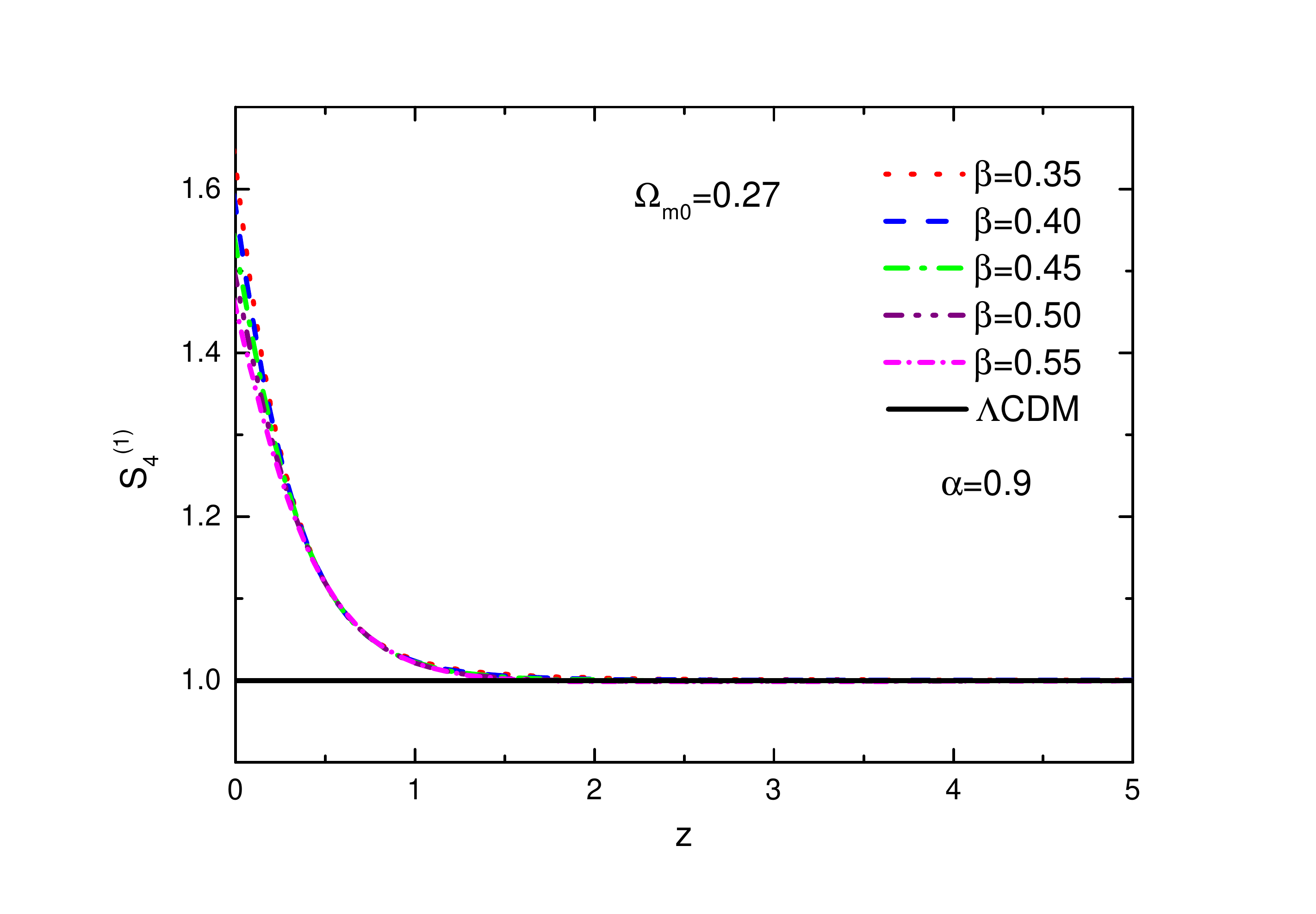}
\caption{(color online). The evolution trajectories of $E$, $q$, $S_3^{(1)}$ and $S_4^{(1)}$, respectively, versus redshift $z$ of ERDE for variable $\beta$ with $\alpha=0.9$, as well compared with the $\Lambda$CDM model marked by a star. The dots represent the present values and the arrows indicate the directions of evolution. Herein $\Omega_{\rm m0}=0.27$.}
\label{fig5}
\end{figure*}

\begin{figure*}[htbp]
\centering
\includegraphics[scale=0.3]{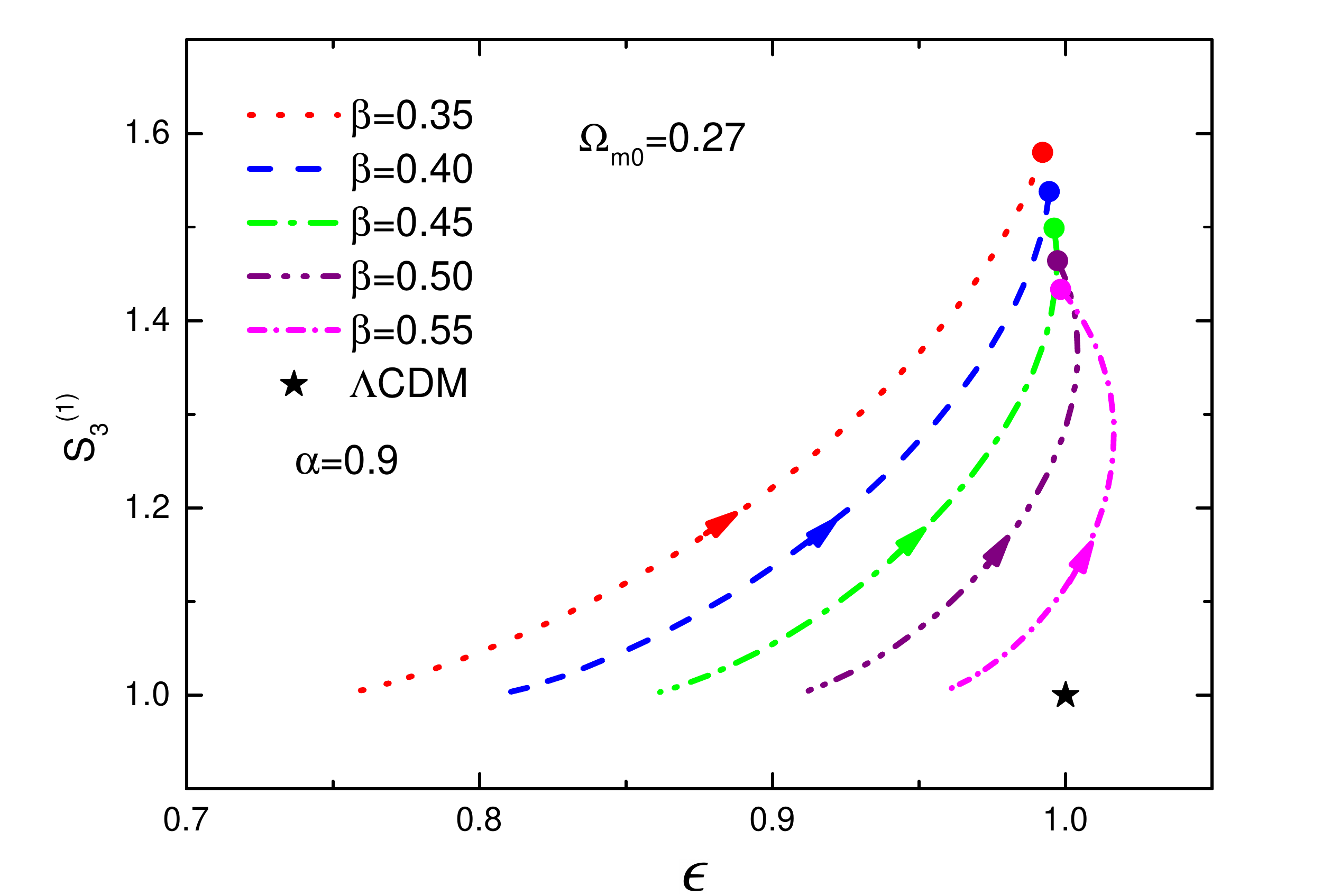}
\includegraphics[scale=0.3]{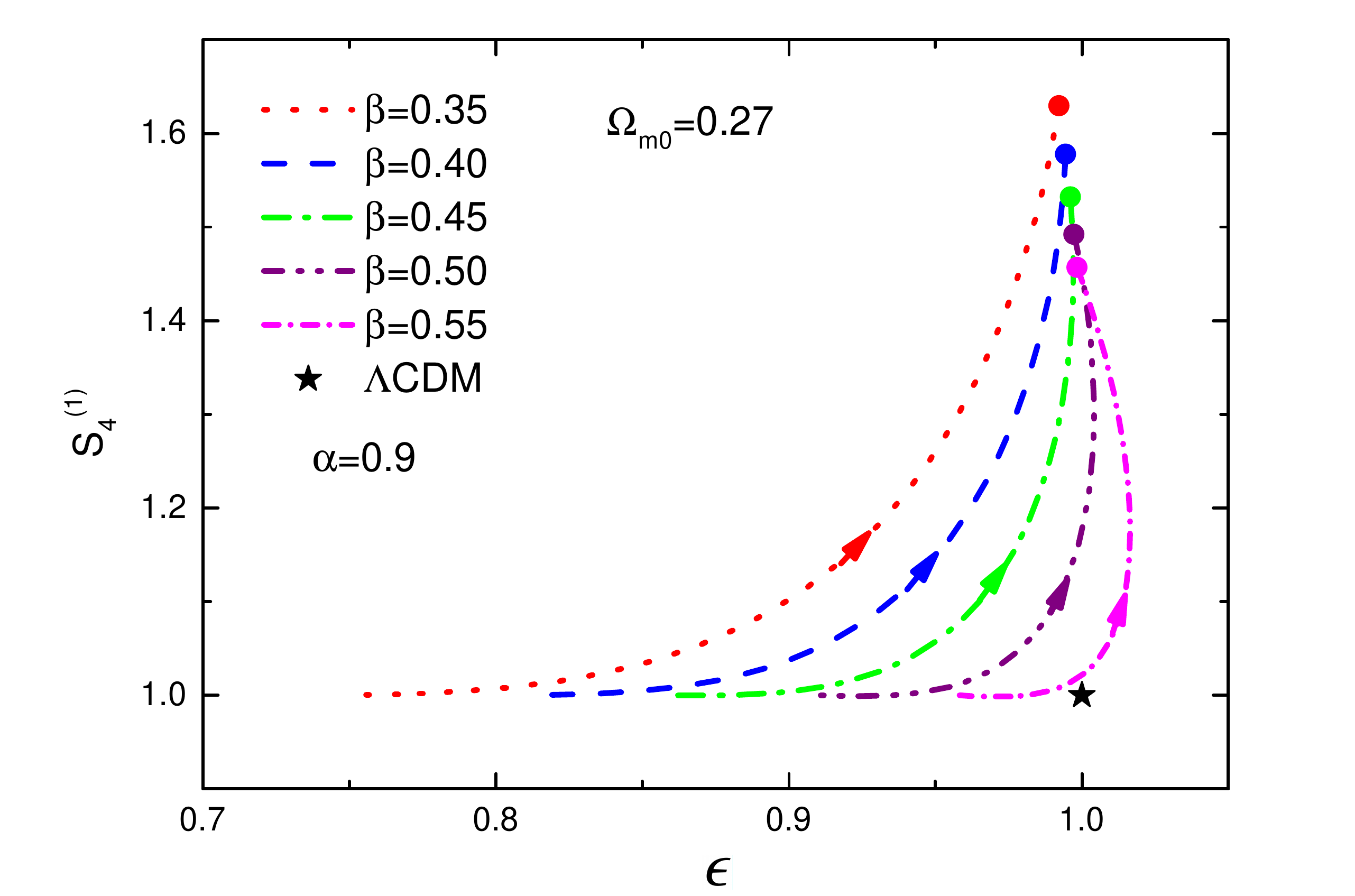}
\caption{(color online). The evolution trajectories of $S_3^{(1)}$ and $S_4^{(1)}$, respectively, versus $\epsilon$ of ERDE for variable $\beta$ with $\alpha=0.9$, as well compared with the $\Lambda$CDM model marked by a star. The dots represent the present values and the arrows indicate the directions of evolution. Herein $\Omega_{\rm m0}=0.27$.}
\label{fig6}
\end{figure*}

At first, it is necessary to clarify the different considerations between our previous work~\cite{YF:2013CTP243} mentioned above and this one, which can help to choose proper typical values of parameters here. In the former, we used an imposed condition $w_0=-1$ to reduce by one the degrees of freedom for parameters $\alpha$ and $\beta$, which were both born arbitrary. But in this paper, from the perspective of simulation in theory, we let $w$ remain free so as to investigate the dependency level of the ERDE model upon $\alpha$ and $\beta$ by adjusting them. Therefore, we explore two cases. The first is adjusting $\alpha$ with a constant $\beta$; the second is, whereas, adjusting $\beta$ with a constant $\alpha$.

For constant $\beta$, we take $\beta=0.5$ as a typical value, approaching the observational constraints~\cite{GCJ:2009PRD043511,WYT:2010PRD083523,Malekjani:2011ASS515}. We fix $\Omega_{\rm m0}=0.27$ throughout the paper. Figure~\ref{fig1} exhibits the phenomenon we mentioned above, namely, ``{\em $\alpha>1$ makes ERDE behave like quintessence while $\alpha<1$ like quintom}''. It is obvious that $\alpha$ plays a key role in the evolution of ERDE. When $\alpha>1$, the EOS evolves in the range of $-1<w<0$. When $\alpha<1$, the EOS evolves from the region of $w>-1$ to that of $w<-1$, i.e., the model exhibits a quintom-like evolution behavior. In particular, the boundary $\alpha=1$ can as well make the model behave like quintessence, but the universe will ultimately enter the de Sitter phase in the far future. It is necessary to emphasize that these features undoubtedly hold for the RDE model~\cite{ZX:2009PRD103509}. In our previous work~\cite{YF:2013CTP243}, although $\beta$ took 0.3, 0.4, 0.5, and 0.6, under the condition of the present EOS value $w_0=-1$, all the values $\alpha$ obtained there were less than 1. Just because of this, with no lack of universality, in this paper we explore the ERDE model comprehensively without missing any possibilities. Thus, around the boundary 1, we take $\alpha$ to be 0.8, 0.9, 1.0, 1.1, and 1.2.

Figure~\ref{fig2} shows the evolutions of the various-order derivatives of the scale factor $a$ versus redshift $z$, from the first to the fourth, for the ERDE model. They are $E$, $q$, $S_3^{(1)}$ and $S_4^{(1)}$, respectively, and they are meanwhile compared with the $\Lambda$CDM model. It can be seen that for $E(z)$, according to Eq.~(\ref{eq7}), in the low-redshift region ($z \lesssim 1$) the curves of the model itself with various parameter values, even together with that of $\Lambda$CDM, are highly degenerate. Although the degeneracy is broken in the high-redshift region, however, the well-known observational data are mainly within the low-redshift region, featuring $z \lesssim 1$. For instance, for some supernova samples~\cite{Conley:2011ApJS1} the majority of the redshifts are in the range of $z<1$ while only a few are in the range of a higher redshift, $1<z<1.4$. Therefore, the current observations for $E(z)$ have not been of help so far. If the next generation Extremely Large Telescopes with high resolution would observe the high-redshift QSOs ($2<z<5$)~\cite{Liske:2008MNRAS1192}, the evolution of $E(z)$ may help effectively.

For $q(z)$, according to Eq.~(\ref{q}), in the low-redshift region the degeneracy that exists in the $E(z)$ case is broken to some extent, but the trends of these curves are quite close to one another, including that for $\Lambda$CDM. Better exhibitions appear in the cases of $S_3^{(1)}$ and $S_4^{(1)}$, especially in the former. From the $S_3^{(1)}(z)$ plot in Fig.~\ref{fig2}, the degeneracy is perfectly broken in the region of $z<1$. The two symmetrical orientations of evolution due to different $\alpha$, which has been concluded before from the $w(z)$ plot of Fig.~\ref{fig1}, appear apparently. When $\alpha>1$, namely $w$ is always larger than $-1$, $S_3^{(1)}$ evolves decreasingly from 1; when $\alpha<1$, namely $w$ can evolve across $-1$, $S_3^{(1)}$ evolves increasingly from 1. But differing from $S_3^{(1)}$, $S_4^{(1)}$ does not show the symmetrical aspect, but only in the same side, although the curves separate well in the low-redshift region. In both plots there is a common feature that when $\alpha=1$, getting $S_3^{(1)}=S_4^{(1)}=1$, the same as that of $\Lambda$CDM. So in the $S_3^{(1)}(z)$ and $S_4^{(1)}(z)$ plots there are two shortcomings. One is a high degeneracy still existing in the high-redshift region. The other is the curves of $\alpha=1$ for ERDE superposing that of $\Lambda$CDM. In face of them, a single diagnostic of geometry fails to be achieved. Instead, we combine it with the fractional growth parameter, namely CND, trying to find a better way.

\begin{table}[tbp]
\centering
\begin{tabular}{|lcccccc|}
\hline
$\alpha$&0.8&0.9&1.0&1.1&1.2&\\
\hline
$\beta$&&&0.5&&&\\
\hline
$S_{30}^{(1)}$&2.088&1.464&1&0.696&0.552&\\
$S_{40}^{(1)}$&2.806&1.492&1&1.043&1.332&\\
$\epsilon_0$&1.0034&0.9974&0.9902&0.9813&0.9702&\\
\hline
$\bigtriangleup S_{30}^{(1)}$&&&1.536&&&\\
$\bigtriangleup S_{40}^{(1)}$&&&1.806&&&\\
$\bigtriangleup\epsilon_0$&&&0.0332&&&\\
\hline
\end{tabular}
\caption{The present values of statefinders and fractional growth parameter, $S_{30}^{(1)}$, $S_{40}^{(1)}$ and $\epsilon_0$, and the differences of them, $\bigtriangleup S_{30}^{(1)}$, $\bigtriangleup S_{40}^{(1)}$ and $\bigtriangleup \epsilon_0$. For each case, $\bigtriangleup S_{30}^{(1)}=S_{30}^{(1)}({\rm max})-S_{30}^{(1)}({\rm min})$, $\bigtriangleup S_{40}^{(1)}=S_{40}^{(1)}({\rm max})-S_{40}^{(1)}({\rm min})$ and $\bigtriangleup \epsilon_0=\epsilon_0({\rm max})-\epsilon_0({\rm min})$.}
\label{table1}
\end{table}

Interestingly, when using CND, $\{S_3^{(1)},\epsilon\}$ and $\{S_4^{(1)},\epsilon\}$ of Fig.~\ref{fig3}, we find that the degeneracy in the high-redshift region can be broken clearly, especially $\{S_3^{(1)},\epsilon\}$ performing far better. To solve the second shortcoming, in both $S_3^{(1)}$-$\epsilon$ and $S_4^{(1)}$-$\epsilon$ plots, ERDE with $\alpha=1$ exhibits a short line segment while $\Lambda$CDM is just a point $\{1,1\}$. We may see the reason at a glimpse of Fig.~\ref{fig4}. That is to say, in the evolution history the fractional growth parameter $\epsilon(z)$ becomes closer and closer to 1 from past to present, but not equal to yet. Since the present values of physical parameters are significant in the research of cosmology, Table~\ref{table1} shows the present values of parameters $S_{30}^{(1)}$, $S_{40}^{(1)}$ and $\epsilon_0$, and the differences of them, for each case $\bigtriangleup S_{30}^{(1)}=S_{30}^{(1)}({\rm max})-S_{30}^{(1)}({\rm min})$ and the same way for $\bigtriangleup S_{40}^{(1)}$ and $\bigtriangleup\epsilon_0$. We can see $\bigtriangleup S_{40}^{(1)}>\bigtriangleup S_{30}^{(1)}$, namely the fourth derivative of $a$, compared with the third derivative, alleviates the degeneracy of present values. But even so, the comparison of either the $S_3^{(1)}$-$z$ and $S_4^{(1)}$-$z$ plots in Fig.~\ref{fig2} or the $S_3^{(1)}$-$\epsilon$ and $S_4^{(1)}$-$\epsilon$ plots in Fig.~\ref{fig3} shows $S_3^{(1)}$ to be performing much better during the evolution process than $S_4^{(1)}$. This indeed violates our habitual judgement: that the higher the order of derivative is, the better the diagnostic performs~\cite{CJL:2014EPJC3100,LJ:2014JCAP12043,YL:150308948,Myrzakulov:2013JCAP10047}.

\begin{table}[tbp]
\centering
\begin{tabular}{|lcccccc|}
\hline
$\alpha$&&&0.9&&&\\
\hline
$\beta$&0.35&0.40&0.45&0.50&0.55&\\
\hline
$S_{30}^{(1)}$&1.580&1.538&1.499&1.464&1.433&\\
$S_{40}^{(1)}$&1.629&1.578&1.532&1.492&1.457&\\
$\epsilon_0$&0.9923&0.9945&0.9961&0.9974&0.9984&\\
\hline
$\bigtriangleup S_{30}^{(1)}$&&&0.147&&&\\
$\bigtriangleup S_{40}^{(1)}$&&&0.173&&&\\
$\bigtriangleup\epsilon_0$&&&0.0061&&&\\
\hline
\end{tabular}
\caption{The present values of statefinders and fractional growth parameter, $S_{30}^{(1)}$, $S_{40}^{(1)}$ and $\epsilon_0$, and the differences of them, $\bigtriangleup S_{30}^{(1)}$, $\bigtriangleup S_{40}^{(1)}$ and $\bigtriangleup \epsilon_0$. For each case, $\bigtriangleup S_{30}^{(1)}=S_{30}^{(1)}({\rm max})-S_{30}^{(1)}({\rm min})$, $\bigtriangleup S_{40}^{(1)}=S_{40}^{(1)}({\rm max})-S_{40}^{(1)}({\rm min})$ and $\bigtriangleup \epsilon_0=\epsilon_0({\rm max})-\epsilon_0({\rm min})$.}
\label{table2}
\end{table}

For constant $\alpha$, although it can be either larger or smaller than 1, which leads to reverse orientations of the evolution, we only take $\alpha=0.9$ as a typical value in this exploration, according to the best-fit values for $\alpha$ from some of the recent constraints~\cite{WYT:2010PRD083523,Malekjani:2011ASS515,Lepe:2010EPJC575}. To obtain feasible evolutions, we take 0.35, 0.4, 0.45, 0.5, and 0.55 for $\beta$. Likewise, we explore the four parameters $E$, $q$, $S_3^{(1)}$ and $S_4^{(1)}$ in Fig.~\ref{fig5} first, as well as make a comparison with the $\Lambda$CDM model. We can see that the ERDE model is insensitive to parameter $\beta$. For the first- and second-order hierarchy $E(z)$ and $q(z)$, high degeneracy appears, even together with $\Lambda$CDM. In the third- and fourth-order cases, the ERDE model itself with various parameter values highly degenerates, but the $\Lambda$CDM model can be discriminated perfectly from ERDE in the low-redshift region. Then let us observe the CND of Fig.~\ref{fig6}. The $S_3^{(1)}$-$\epsilon$ and $S_4^{(1)}$-$\epsilon$ curves look as good as the above case of $\beta=0.5$ in Fig.~\ref{fig3}. In both plots the evolution trajectories separate quite well, but the combination of $\{S_3^{(1)},\epsilon\}$ is slightly better than $\{S_4^{(1)},\epsilon\}$ because of the more legible separation in-between curves in the high-redshift region. In the same way we show in Table~\ref{table2} the present values of $S_{30}^{(1)}$, $S_{40}^{(1)}$ and $\epsilon_0$, and the differences of them for $\alpha=0.9$. Likewise the relation $\bigtriangleup S_{40}^{(1)}>\bigtriangleup S_{30}^{(1)}$ demonstrates once again that the fourth-order hierarchy can help to alleviate the degeneracy of present values when compared with the third one. But for the same reason as of the comparison of $S_3^{(1)}$-$\epsilon$ and $S_4^{(1)}$-$\epsilon$ plots in Fig.~\ref{fig6}, we find for the ERDE model $\{S_3^{(1)},\epsilon\}$ is a more efficient parameter pair of diagnostic than $\{S_4^{(1)},\epsilon\}$, which is already concluded in the above-mentioned case of $\beta=0.5$.

\section{Conclusion}
In this paper, we explore the extended Ricci dark energy model with statefinder hierarchy supplemented by the growth rate of perturbations. Since in ERDE there are two independent variables $\alpha$ and $\beta$, we just adjust them, respectively, leaving other parameters fixed, for the sake of investigating the effects of $\alpha$ and $\beta$ on this model. First, a feature of the holographic Ricci-type dark energy models is corroborated again, namely, $\alpha>1$ makes them behave like quintessence while $\alpha<1$ like quintom. For the ERDE model with $\beta=0.5$, letting $\alpha$ vary around 1, we conclude that the evolutions of the Hubble expansion rate $E$ are in high degeneracy in the low-redshift region of $z \lesssim 1$; but because the observational data come mainly from the low-redshift region $z \lesssim 1$, the broken degeneracy in the high-redshift makes no sense. The evolutions of deceleration parameter $q$ do degenerate no more in the low-redshift region of $z \lesssim 1$. However, for both $E$ and $q$ plots, the evolution of $\Lambda$CDM cannot be singled out from in-between with great ease. The situations of $S_3^{(1)}$ and $S_4^{(1)}$, which contain the third and fourth derivatives of the scale factor, respectively, turn out to be better. $S_3^{(1)}(z)$ evolves with respect to redshift $z$ along two orientations symmetrical to each other on the basis of different $\alpha$ in the region of $z<1$. When $\alpha>1$, it evolves decreasingly from 1; when $\alpha<1$, it evolves increasingly from 1. $S_4^{(1)}$ although seems featureless by contrast with $S_3^{(1)}(z)$, by a comparison of the present-value differences of the parameters $S_{3}^{(1)}$ and $S_{4}^{(1)}$ ($\bigtriangleup S_{30}^{(1)}=1.536$, $\bigtriangleup S_{40}^{(1)}=1.806$), we see that $S_4^{(1)}$ is capable of alleviating the degeneracy existing in other statefinder parameters for ERDE. There are also two unsolved problems that high degeneracy still exists in the high-redshift region, and with $\alpha=1$ are degenerate with $\Lambda$CDM. As for them, the combination of statefinder hierarchy $S_n$ and fractional growth parameter $\epsilon$ (CND) can help. In both $S_3^{(1)}$-$\epsilon$ and $S_4^{(1)}$-$\epsilon$ plots, the degeneracy in the high-redshift region is pretty broken and ERDE with $\alpha=1$ exhibits a short line segment, but $\Lambda$CDM exhibits just a point $\{1,1\}$. Nevertheless, the fact that the $S_3^{(1)}$-$z$ and $S_3^{(1)}$-$\epsilon$ planes feature a more legible and regular sight of evolution due to $\alpha$ than that of the $S_4^{(1)}$-$z$ and $S_4^{(1)}$-$\epsilon$ planes, reveals that the third-order statefinder hierarchy $S_3^{(1)}$ makes more sense for ERDE than $S_4^{(1)}$ does.

For the ERDE model with $\alpha=0.9$, we find by contrast with $\alpha$, $\beta$ has a weak influence. The evolution trajectories of $E$, $q$, $S_3^{(1)}$ and $S_4^{(1)}$ with respect to redshift $z$ are in high degeneracy within the ERDE model. The degeneracy of ERDE with $\Lambda$CDM still exists for $E(z)$ and $q(z)$, but it is perfectly broken in the low-redshift region for both $S_{3}^{(1)}(z)$ and $S_{4}^{(1)}(z)$. As for the high-redshift region, the use of CND can break the degeneracy there, especially the $\{S_3^{(1)},\epsilon\}$ pair performs more efficiently than the $\{S_4^{(1)},\epsilon\}$ pair, although $\bigtriangleup S_{30}^{(1)}=0.147<\bigtriangleup S_{40}^{(1)}=0.173$. $\Lambda$CDM can also be discriminated from ERDE by CND. As a consequence of all the materials studied above, we find that, although the higher-order statefinder hierarchy, even with the growth rate of perturbations, can differentiate the ERDE model itself with various parameter values and also from the $\Lambda$CDM model, there is the interesting discovery that the third-order hierarchy of statefinder is really a better choice than the fourth-order hierarchy for the ERDE model.

\begin{acknowledgments}
This work was supported by the National Natural Science Foundation of
China under Grant No.~11175042, the Provincial Department of Education of
Liaoning under Grant No.~L2012087, and the Fundamental Research Funds for the
Central Universities under Grants No.~N140505002, No.~N140506002, and No.~N140504007.
\end{acknowledgments}

\end{document}